\documentclass[showpacs,prl,twocolumn,,groupedaddress]{revtex4}
\usepackage{epsfig}
\usepackage{bm}

\begin{document}

\title{Dimer phases in quantum antiferromagnets with orbital degeneracy}
\author{G.~Jackeli}
\altaffiliation[]{george.jackeli@epfl.ch}
\author{D.~A.~Ivanov}
\affiliation{Institute of Theoretical Physics,
Ecole Polytechnique F\'ed\'erale de Lausanne, CH-1015 Lausanne,
Switzerland}

\date{\today}
\begin{abstract}
We study and solve the ground-state problem of a microscopic model for a family of orbitally degenerate
quantum magnets. The orbital degrees of freedom are assumed to have directional character and are
represented by static Potts-like variables. In the limit of vanishing Hund's coupling,
the ground-state manifold of such a model is spanned by the hard-core dimer (spin singlet)
coverings of the lattice. The extensive degeneracy of dimer coverings is lifted at a
finite Hund's coupling through an order-out-of-disorder mechanism by virtual triplet excitations.
The relevance of our results to several experimentally studied systems is  discussed.
\end{abstract}
\maketitle

Isotropic quantum spin systems in dimensions higher than one  have typically magnetically long-range
ordered ground states and thus gapless spin excitations. This  picture may, however, fail for so-called
frustrated antiferromagnets, which have extensively degenerate classical ground states \cite{frusrev}.
In such systems exotic quantum phases without long-range magnetic order can emerge as the true ground states.
Spin systems can be frustrated by the geometry of the lattice or by the presence of competing interactions.
In this letter, we discuss another scenario in which frustration of spin interactions is
induced by the presence of additional electronic degrees of freedom, such as orbital degeneracy.
A possibility of formation of orbitally driven  magnetically disordered states
has been suggested within various coupled spin-orbital models \cite{pen,feiner,Timy1,mila}.

Here we consider a generalization of the spin-orbital model of Ref.~\cite{Timy1} to a wide class
of lattices in arbitrary dimensions. We study and solve the ground-state problem for this model.
We find remarkably that orbital degeneracy drives the spin degrees of freedom into an extensively
degenerate manifold of spontaneously dimerized states. We demonstrate that various types of valence
bond crystal (VBC) states can be selected by an order-out-of-disorder mechanism due to virtual
triplet excitations. We motivate our study by several experimentally discovered
spin gapped systems and discuss the relevance of our results to these materials.

{\it The model}.--
The model is defined on an arbitrary lattice with links aligned along $z$ distinct directions.
We assume that at each lattice site there are $z$ orbitally degenerate orthogonal levels such that
each of them has hopping amplitudes only along the corresponding direction \cite{note1}. We further
assume that, due to the strong on-site repulsion $U$, the system is in the Mott insulating state with
exactly one electron per lattice site. The on-site repulsion also contains a small Hund's coupling $J_{\rm H}$
penalizing virtual processes with spin-zero state of a doubly occupied site. Such a model has
experimental realizations on different lattices (see our discussion in the corresponding section).
To the lowest order in $t/U$, where $t$ is the hopping amplitude (assumed to be the same
for all links and corresponding orbitals), 
it leads to an effective Kugel-Khomskii-type Hamiltonian \cite{KK}.
Since second-order virtual processes locally conserve orbital flavors,
the orbital degrees are static Potts-like variables.
The effective spin-orbital Hamiltonian for such a system takes the form \cite{Timy1}:
\begin{eqnarray}
H= J \sum_{\langle ij\rangle}{\Big \{} {\big [} \vec S_i\cdot \vec S_{j}-\frac{1}{4} {\big ]}O_{ij}
-{\big [} \zeta + \eta \vec S_i\cdot \vec S_{j}{\big ]} {\tilde O}_{ij}{\Big \}}
\label{eq1}
\end{eqnarray}
where the sum is taken over pairs of nearest-neighbor (NN) sites. To the leading order in $J_{\rm H}/U$,
the coupling constants are given by $J\approx4t^2/U$, $\eta \approx J_{\rm H}/(2U)$, and
$\zeta \approx \frac{1}{4}$.  The small dimensionless parameter
$\eta$ characterizes the effective strength of the Hund's coupling. The full expressions
for the coupling constants in terms of $t$, $U$, and $J_{\rm H}$ can be easily obtained from the exchange energies given in Ref.~\onlinecite{Timy1}.
For our further analysis, it will be important that $\zeta\ge \frac{1}{4}$ for all values
of the Hund's coupling.

The first term of the Hamiltonian (\ref{eq1}) describes the antiferromagnetic (AF) coupling between
NN spins and is active when both ends of the link are occupied by the orbitals corresponding to the
link direction. If $\alpha_{ij}$ denotes the ``color'' corresponding to the direction of the link $ij$,
the operator $O_{ij}$ is defined as  $O_{ij}=\delta_{\sigma_{i},\alpha_{ij}}\delta_{\sigma_{j},\alpha_{ij}}$
where $\delta$ is the Kronecker symbol, and $\sigma_{i}=1,...,z$ is the Potts-like variable for orbital flavor
at the site $i$. The second term corresponds to the situation when on the bond $ij$ only one orbital
is of the color $\alpha_{ij}$. The corresponding projector operator is
${\tilde O}_{ij}=\delta_{\sigma_{i},\alpha_{ij}}[1-\delta_{\sigma_{j},\alpha_{ij}}]
+\delta_{\sigma_{j},\alpha_{ij}}[1-\delta_{\sigma_{i},\alpha_{ij}}]$.

For further analysis it is convenient to rewrite the Hamiltonian as the sum of three terms (in the units of $J$):
\begin{eqnarray}
H \!\! &=& \!\! H_{\rm AF} + H_{\rm FM} + E_0\, , \nonumber\\
H_{\rm AF} \!\! &=& \!\! \sum_{\langle ij\rangle}{\big [} \vec S_i\cdot \vec S_{j}+
\big ( 2\zeta-\frac{1}{4} \big ) {\big ]}O_{ij}\, , \\
H_{\rm FM} \!\! &=& \!\! - \eta  \sum_{\langle ij\rangle} \vec S_i\cdot \vec S_{j} {\tilde O}_{ij}\, ,
\nonumber
\end{eqnarray}
and $E_0$ is a static energy depending on the on-site ``color'' variables, but not on the dynamic spin
degrees of freedom. It counts the number of matches between the colors of the link and one of its ends:
$E_0=-\zeta \sum_{i,j} \delta_{i,\alpha_{ij}}$.

This model can be solved exactly in the case of zero Hund's coupling
$\eta=0$ on most commonly considered lattices (see the following
section), and a further perturbative analysis at finite $\eta$ is possible on many
lattices, including the triangular and square lattice (see the corresponding section).

{\it Zero Hund's coupling.}--
We start our analysis from the limit of zero Hund's coupling  $\eta=0$. In this limit, we show
that the ground state is extensively degenerate and is described by dimer coverings of the lattice.

\begin{figure}
\epsfysize=23mm
\centerline{\epsffile{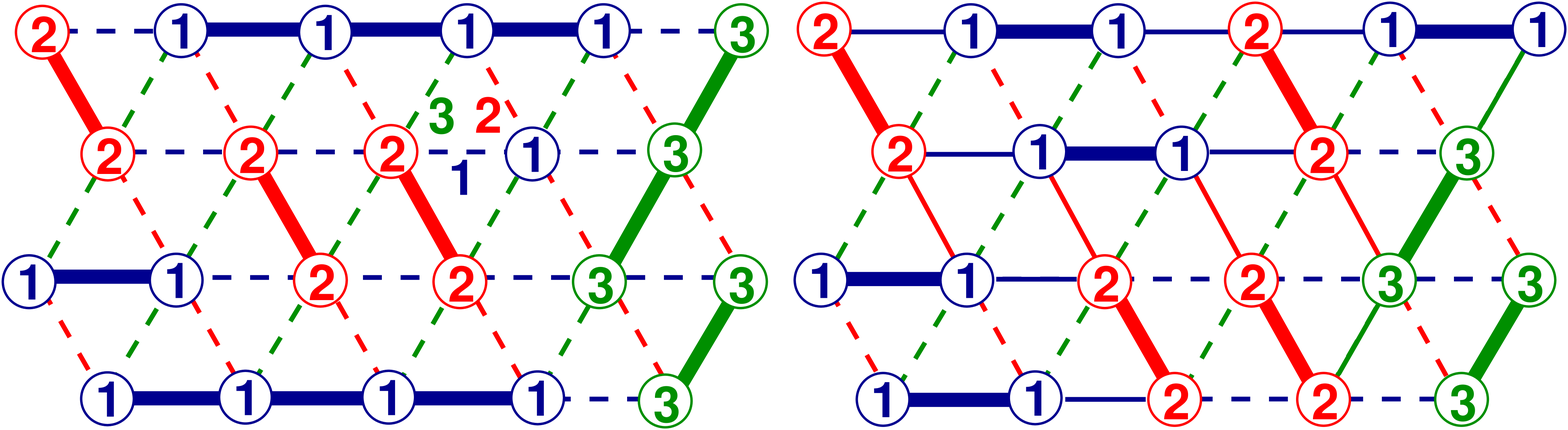}}
\caption{(Color online)
Left:
A possible orbital pattern and the corresponding decoupled AF chains on the triangular lattice,
 in the limit of zero Hund's coupling.
The numbers label the three bond directions and the occupied orbitals on sites.
Right:
An example of the dimer covering minimizing the energy at zero Hund's coupling.
Thick (thin) solid lines denote  AF (FM) intra-(inter-)dimer bonds, respectively.}
\label{fig:oo}
\end{figure}

The term $E_0$ in the Hamiltonian only depends on the static orbital variables (colors).
The antiferromagnetic term $H_{\rm AF}$ is active only on the links whose colors match those at
both their ends. The active AF bonds thus form non-intersecting linear chains (Fig.~\ref{fig:oo}, left).
The longer is the chain, the more energy can be gained from the AF spin interaction.
However, for each AF link we pay a positive energy $(2\zeta-\frac{1}{4}) \ge 1/4$
(the second term in $H_{\rm AF}$).
We can prove that under these conditions the minimal possible AF energy is achieved when all
AF chains are dimers. A proof follows from the variational estimate on the ground-state energy
of the $M$-site AF Heisenberg chain with open ends: $E_M\ge \frac{1}{4}-\frac{M}{2}$, with the equality attained
only at $M=2$. [This estimate is, in turn, obtained by dividing the chain into shorter overlapping
sub-chains of lengths two and three with exactly known energies $E_2=-3/4$ and $E_3=-1$.]

Therefore, at zero Hund's coupling  $\eta=0$, the ground-state coloring of vertices corresponds
to a dimer covering of the lattice (Fig.~\ref{fig:oo}, right), provided that this dimer covering simultaneously minimizes
$E_0$. The minimization of $E_0$ is, for example, always possible for lattices where each site
has equal valencies for all colors of outgoing links. This condition is satisfied for most
well-known lattice types: triangular, square (cubic), kagom\'e, pyrochlore, checkerboard, hexagonal,
etc. In addition, dimer coverings minimizing $E_0$ may also exist for other exotic lattices
not satisfying the equal-color-valency constraint.

The dimer coverings corresponding to the ground state must obey an additional ``no-chain''
constraint: no two neighboring dimers can lie on the same line (otherwise they form a longer
AF chain which is energetically unfavorable). Even with this constraint, the entropy of such
dimer coverings remains extensive for most commonly considered lattices. Thus we conclude that,
on such lattices, the model is in the dimer-liquid state at zero Hund's coupling.

{\it Order out of disorder by triplet fluctuations}.--
We now introduce a finite value of Hund's coupling $\eta>0$ and show that it
lifts the extensive degeneracy of dimer coverings by triplet fluctuations.
The Hund's coupling produces a ferromagnetic term $H_{\rm FM}$ active only on
links with exactly one end matching the link color. Given a dimer covering
from the previous consideration, the FM bonds correspond to the ``legs'' of dimers:
adjacent links along the dimer direction (Fig.~\ref{fig:oo}, right). We assume that
the Hund's coupling is small $\eta\ll 1$ and treat it perturbatively. A convenient
method to build a perturbative expansion is to rewrite the FM coupling between
two ends of dimers $i$ and $j$ in the basis of triplet excitations on those
dimers \cite{SB}. At each dimer, we introduce three triplet excitations
$t_{i,\alpha}^{\dag}$ characterized by the spin polarization
$\alpha=x,y,z$. Those triplet excitations may be treated as hard-core bosons,
and the AF part of the Hamiltonian is written (in the units of $J$) as
$H_{\rm AF}= \sum_{i,\alpha} t_{i,\alpha}^{\dag} t_{i,\alpha}$.
The FM part of the Hamiltonian $H_{\rm FM}$ is treated as a perturbation and
is written as the sum of the three contributions
\begin{eqnarray}
H_{1}(i,j)&=&-2\lambda m_{i}m_{j}\Bigl[t_{i,\alpha}^{\dag}t_{j,\alpha}^{\dag}+t_{i,\alpha}^{\dag}t_{j,\alpha}
+h.c.\Bigr]\\
H_{2}(i,j)&=&2\lambda\Bigl[i\epsilon_{\alpha\beta\gamma}m_{j}t_{j,\alpha}^{\dag}t_{i,\beta}^{\dag}t_{i,\gamma}+[i\leftrightarrow j]+h.c.\Bigr]\nonumber\\
H_{3}(i,j)&=&2\lambda\epsilon_{\alpha\beta\gamma}\epsilon_{\alpha\beta^{\prime}\gamma^{\prime}}t_{i,\beta}^{\dag}t_{i,\gamma}
t_{j,\beta^{\prime}}^{\dag}t_{i,\gamma^{\prime}},
\nonumber
\label{eq3}
\end{eqnarray}
where $m_{i}=\pm$ is a phase factor depending on the orientation of the singlet wave function on the dimer \cite{SB}.
We also introduce $\lambda=\eta/8$, to simplify combinatoric coefficients.

\begin{figure}
\epsfysize=33mm
\centerline{\epsffile{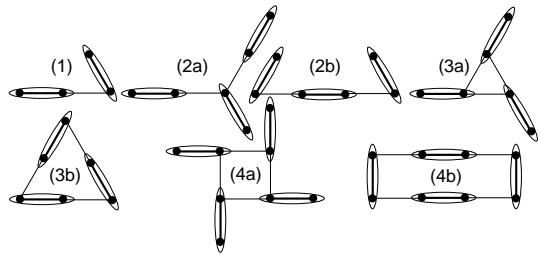}}
\caption{Examples of topologically nonequivalent linked clusters.}
\label{fig:graphs}
\end{figure}

A convenient way to analyze perturbative corrections is provided by the linked-cluster expansion \cite{GS}.
The weak inter-dimer coupling $H_{\rm FM}$ is treated as a perturbation, and
the correction to the ground-state energy may be written as a sum over
all distinct linked clusters $\Delta E=\sum_{\cal C}N_{\cal C}E_{\cal C}$.
Here $N_{\cal C}$ is the number of ways the cluster ${\cal C}$ can be embedded in the lattice
with a given dimer covering and $E_{\cal C}$ is the perturbative contribution to the energy from the cluster ${\cal C}$
(given by a power series in $\lambda$). Examples of topologically distinct clusters are presented in        
Fig.~\ref{fig:graphs}. Note that the leading contribution to $E_{\cal C}$ from tree-like clusters
with $n$ FM links is of order $\lambda^{2n}$ [the clusters $(1)$, $(2a)$, and $(2b)$ in Fig.~\ref{fig:graphs}],
while loop clusters with $n$ links [the clusters $(3a)$, $(3b)$, $(4a)$ and $(4b)$] contribute to the order $\lambda^{n}$.
The actual calculation of $E_{\cal C}$ involves lengthy combinatorial analysis of possible virtual processes and will be
reported elsewhere.
Here we just present the results of those calculations in application to the triangular and square lattices.

The leading perturbative contribution comes from the cluster $(1)$ and is given by $E_{(1)}=-6\lambda^2(1-2\lambda + \lambda^2 + \dots)$.
However, the total number of such clusters is independent of the dimer covering
(it is equal to the number of inter-dimer bonds, {\it i.e.}, twice the number of dimers), therefore
the degeneracy is not lifted to this order.

\begin{figure}
\epsfysize=29mm
\centerline{\epsffile{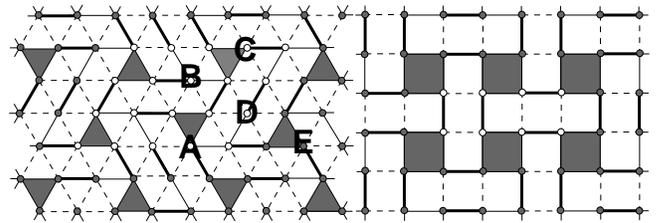}}
\caption{Left: The ground state dimer covering of a triangular lattice.
The triangular loops are explicitly shown as dark triangles.
A possible choice of an unit cell is shown by open circles.
Right: The ground state dimer covering of a square lattice.
The length-four loops are shown as dark squares and white rectangles.}
\label{fig:vbc}
\end{figure}

On the triangular lattice, the degeneracy is lifted to the leading order by the contributions from the cluster
types $(3a)$ and $(3b)$ in Fig.~\ref{fig:graphs}. The cluster $(3a)$ generates a non-frustrated FM coupling while
the cluster $(3b)$ is frustrated. The contributions of these two clusters differ in sign, $E_{(3a)}=-E_{(3b)}=-36\lambda^3$,
to the leading order in $\lambda$. Thus, on the triangular lattice, the ground state should maximize the number
of $(3a)$ type clusters and minimize the number of $(3b)$ type clusters. We prove that such an optimization
leads to a dimer crystal shown in Fig.~\ref{fig:vbc} (left panel). It provides the maximal density $N/5$ ($N$ is the
number of sites) of $(3a)$ type clusters and no $(3b)$ clusters. The unit cell of such a crystal contains 20 sites,
and its degeneracy is 60-fold. A proof of crystallization may be performed by
dividing the triangular lattice into edge-sharing hexagons composed of six triangles each.
On a hexagon, we can place at most one triangular loop of a cluster $(3a)$.
A hexagon which has one of its edges shared by a triangular loop
placed on an adjacent hexagon can not itself host a triangular loop.
An empty hexagon can have at most two such edges (furthermore, they must be non-parallel and disconnected).
To achieve the maximal density of triangular loops, we start with a three-hexagon cluster composed of an empty
hexagon {\bf B} sharing two edges with the triangular loops on the adjacent hexagons {\bf A} and {\bf C}
(Fig.~\ref{fig:vbc}, the letters mark the centers of hexagons). Then the hexagon {\bf D} must also be empty
and, moreover, it may have at most one edge shared by a triangular loop (on the hexagon {\bf E}).
One can show that the maximal density of $(3a)$ clusters is achieved by a close-packed covering of the
triangular lattice by five-hexagon clusters of the type {\bf A}-{\bf E}. Now starting from one such
cluster and choosing different possible divisions of the lattice into hexagons one can verify that the
corresponding dimer covering is unique and periodic, as shown in Fig.~\ref{fig:vbc} (left panel).

On the square lattice, a similar analysis needs to be done to the order $\lambda^4$, since length-three
loop clusters are not possible. At this order, the contributions come from the length-two clusters $(2a)$ and
$(2b)$ and from the length-four loops (such as $(4a)$ and $(4b)$), see Fig.~\ref{fig:graphs} (other length-four
loop clusters are also possible, but not shown).
The  clusters $(2a)$ and $(2b)$  have different contributions:
$E_{(2a)}=12\lambda^4$ and $E_{(2b)}=-20\lambda^4$. All length-four loop clusters involve unfrustrated
FM couplings and contribute the same energy $E_{(4)}=-120\lambda^4$.
With those cluster energies, we can prove that the resulting dimer covering forms a crystal
shown in Fig.~\ref{fig:vbc} (right panel). The proof is based on the observation that the
energy gain from length-four loop clusters is much larger than the scale of energy optimization between
$(2a)$ and $(2b)$ type clusters. Therefore, an optimal configuration should involve as many length-four
loops, as possible (each FM link should belong to two such loops). Under this condition, an optimization
of the energy contribution from clusters $(2a)$ and $(2b)$ leads to the crystal
structure shown in the figure. Its unit cell contains 8 lattice sites, and the degeneracy is 8-fold.

On other lattices, we expect that virtual triplet processes also lead to a dimer
crystallization at some order of the perturbation theory. A similar analysis may be performed
individually for a given lattice.

{\it Experimental systems}.-- The considered model and the derived results may be relevant for
several experimentally studied materials.

In the layered compound NaTiO$_2$ \cite{natio2n,natio2a} the spin one-half Ti$^{3+}$ ions form a triangular lattice
and have one $d$-electron in the three-fold degenerate $t_{2g}$ manifold ($d_{xy}$, $d_{xz}$, and $d_{yz}$ orbitals).
The three non-equivalent bonds of the triangular lattice are aligned along the $xy$, $xz$, and $yz$ directions, and  
each orbital state has a dominant hopping amplitude along the corresponding bond.
This system does not show any signature of long-range magnetic order at low temperatures \cite{natio2n}.
A drop of magnetic susceptibility has been observed \cite{natio2a}, but no dimerized superstructure has yet been detected.

The compound Sr$_{2}$VO$_4$  is a possible realization of the model (\ref{eq1})
on the square lattice (formed by spin one-half Vanadium ions).
Each V$^{4+}$ ion is surrounded by the distorted oxygen octahedra elongated in the $z$-direction \cite{sr2vo4a}
and single $d$ electron occupies a two-fold degenerate ($d_{xz}$ and $d_{yz}$) level.
On the square lattice in the $xy$ plane, the $d_{xz}$ and $d_{yz}$ electrons hop
only in the $x$ and $y$ directions, respectively.
The experimental study of Sr$_{2}$VO$_{4-\delta}$ \cite{sr2vo4b} suggested an absence of
magnetically ordered states down to zero temperature
and a possible formation of singlet pairs in the stoichiometric compound $\delta=0$.
The available experimental data are, however, inconclusive about the presence of
a spin gap in this compound.

The model (\ref{eq1}) on the pyrochlore lattice describes the
MgTi$_2$O$_4$ spinel compound ~\cite{Timy1}. The latter has a spin gap
and a dimerized bond pattern  at low temperatures \cite{mgti2o4}.
It has been shown in Ref.~\cite{Timy1}, within the mean-field like approach, that the dimer states for
model (\ref{eq1}) on a pyrochlore lattice are favorable over some other magnetic states.
Here we provide a proof that dimer states are the ground states.
Our mechanism of the selection of VBC pattern would, however, predict a different dimer pattern
than observed experimentally in MgTi$_2$O$_4$.
This suggests that in this compound  the degeneracy is lifted by a magnetoelastic mechanism, as discussed
in Ref.~\cite{Timy1}, and not by quantum fluctuations. An alternative mechanism for observed
dimerized pattern within a different (itinerant-electron) model has recently been suggested in Ref.~\cite{KM}.  

Note that our model has two different energy scales. At the higher
energy scale (of order $J$), the system organizes into classically
fluctuating singlets (dimers). This possibly explains the spin gap in the
compounds mentioned above. The crystallization of dimers by
triplet fluctuations occurs at a much lower energy scale (determined by
the $\eta J$ term in Eq.~\ref{eq1}). In real systems, this is one of several
possible mechanisms of crystallization (including magnetoelastic effects,
residual overlaps between orbitals in non-dominant directions, etc.)
For any particular compound, a theoretical prediction of the crystallization
pattern would require a thorough analysis of different ordering mechanisms.
On the experimental side, an identification of the dimerization pattern
may be difficult in the case of large unit cells like those predicted
in this work.

{\it Summary}.-- To summarize, we have studied a spin-orbital model
for a family of quantum antiferromagnets with orbital degeneracy.
In such systems, the orbital degrees of freedom induce a spontaneous
dimerization of the spins. The resulting extensively degenerate
ground state is equivalent to a problem of constrained classical hard-core dimers.
At a lower energy scale, the extensive degeneracy is lifted through an
order-out-of-disorder mechanism by virtual triplet excitations
and a valence-bond crystal is formed. At intermediate temperatures,
above the dimer crystallization,
the spin system is in a liquid state. Our study proposes a mechanism
of the spin-gap formation experimentally observed in several two- and
three-dimensional systems. An interesting possible extension of our
model may include additional terms arising
from neglected overlaps of orbitals and
leading to a quantum dynamics of dimers. Such an extension might be
a better candidate than pure spin systems for
a realization of quantum dimer models \cite{RK,MS} and of the long-sought
quantum spin-liquid phase.

We are grateful to  A. G . Abanov, A. Honecker, D. I. Khomskii, C. Mudry, B. Normand and M. E. Zhitomirsky for
interesting and inspiring discussions. G.J. acknowledges support by GNSF under the Grant No.06-81-4-100.



\begin{thebibliography}{20}
\bibitem{frusrev} For reviews, see A.P. Ramirez in {\it Handbook of Magnetic Materials},
 edited by K. H. J. Buschow (North-Holland, Amsterdam, 2001);
R. Moessner, Can. J. Phys. {\bf 79}, 1283 (2001);
 G. Misguich and C. Lhuillier, cond-mat/0310405,
{\it Magnetic systems with competing interactions}, edited by H.T. Diep
 (World Scientific, Singapore, 2005).
\bibitem{pen}
H. F. Pen, J. van den Brink, D. I. Khomskii, and G. A. Sawatzky,  Phys. Rev. Lett. {\bf 78}, 1323 (1997).
\bibitem{feiner}
L.F. Feiner, A.M. Oles, and J. Zaanen, Phys. Rev. Lett. {\bf 78}, 2799 (1997).
\bibitem{Timy1} S. Di Matteo, G. Jackeli, C. Lacroix, and N. B. Perkins, Phys. Rev. Lett. {\bf 93}, 077208 (2004);
S. Di Matteo, G. Jackeli, and N. B. Perkins, Phys. Rev. B {\bf 72}, 024431 (2005).
\bibitem{mila} F. Vernay, K. Penc, P. Fazekas, and F. Mila,
Phys. Rev. B 70, 014428 (2004).
\bibitem{note1}
For certain lattices, several directions may be treated as equivalent, e.g., two diagonal directions
in the checkerboard lattice or pairs of perpendicular directions in the pyrochlore lattice.
We only require that such equivalent directions never intersect at any lattice site.
\bibitem{KK}  K. I. Kugel and D. I. Khomskii,
Usp. Fiz. Nauk {\bf 136}, 621 (1982) [Sov. Phys. Usp. {\bf 231}, 25 (1982)].
\bibitem{SB} S. Sachdev and R. N. Bhatt, Phys. Rev. B {\bf 41}, 9323 (1990).
\bibitem{GS}M. P. Gelfand and R. R. P. Singh, Adv. Phys. {\bf 49}, 93 (2000).  
\bibitem{natio2n} K. Hirakawa, H. Kadowaki, and K. Ubukoshi, J. Phys. Soc. Jpn. {\bf 54}, 3526 (1985).
\bibitem{natio2a} K. Takeda {\it et al\/}., J. Phys. Soc. Jpn. {\bf 61}, 2156 (1992).
\bibitem{sr2vo4a} M. Itoh {\it et al\/}., Solid State Commun. {\bf 80}, 545 (1991)
\bibitem{sr2vo4b} N. Suzuki, T. Noritake, and T. Hioki, Mat. Res. Bull. {\bf 27}, 1171 (1992).
\bibitem{mgti2o4} M. Isobe and Y. Ueda, J. Phys. Soc. Jpn. {\bf 71}, 1848 (2002);
M. Schmidt {\it et al\/}., Phys. Rev. Lett. {\bf 92}, 056402 (2004).
\bibitem{KM}  D. I. Khomskii and T. Mizokawa, Phys. Rev. Lett. {\bf 94}, 156402 (2005).
\bibitem{RK}  D. S. Rokhsar and S. A. Kivelson, Phys. Rev. Lett. {\bf 61}, 2376 (1988).
\bibitem{MS}   R. Moessner and S. L. Sondhi, Phys. Rev. Lett. {\bf 86}, 1881 (2001)
\end{thebibliography}
\end{document}